\documentclass{jfm}

\usepackage{graphicx}
\usepackage{dcolumn}
\usepackage{bm}
\usepackage{amssymb}
\usepackage{CJK}
\usepackage{amsmath,mathtools}
\usepackage{color}
\usepackage{hyperref}
\usepackage{psfrag}
\usepackage{stmaryrd}
\usepackage{amssymb}
\usepackage{wasysym}
\usepackage[latin1]{inputenc}
\usepackage{booktabs}
\usepackage{multirow}
\usepackage{natbib}

\shorttitle{Turbulent drag in a rotating frame}
\shortauthor{A. Campagne, N. Machicoane, B. Gallet, P.-P. Cortet
and F. Moisy}

\title{Turbulent drag in a rotating frame}

\author{Antoine Campagne\aff{1}, Nathana\"el Machicoane\aff{1}, Basile Gallet\aff{2}, Pierre-Philippe Cortet\aff{1}, \and Fr\'{e}d\'{e}ric Moisy\aff{1}}

\affiliation{\aff{1}Laboratoire FAST, CNRS, Univ. Paris-Sud,
Université Paris-Saclay, 91405 Orsay, France \aff{2}Service de
Physique de l'\'Etat Condens\'e, CEA, CNRS, Université
Paris-Saclay, CEA Saclay, 91191 Gif-sur-Yvette, France}

\begin{document}

\maketitle

\begin{abstract}

What is the turbulent drag force experienced by an object moving
in a rotating fluid? This open and fundamental question can be
addressed by measuring the torque needed to drive an impeller at
constant angular velocity $\omega$ in a water tank mounted on a
platform rotating at a rate $\Omega$. We report a dramatic
reduction in drag as $\Omega$ increases, down to values as low as
$12$\% of the non-rotating drag. At small Rossby number $Ro =
\omega/\Omega$, the decrease in drag coefficient $K$ follows the
approximate scaling law $K \sim Ro$, which is predicted in the
framework of nonlinear inertial wave interactions and
weak-turbulence theory. However, stereoscopic particle image
velocimetry measurements indicate that this drag reduction rather
originates from a weakening of the turbulence intensity in line
with the two-dimensionalization of the large-scale flow.

\end{abstract}

\section{Introduction}

Determining the drag force on a moving object is a central
question of turbulence research and the main goal of aerodynamics.
A characteristic feature of turbulent flows is the ``dissipation
anomaly'': the drag force becomes independent of the fluid's
viscosity when the latter is low enough~\citep{Frisch}. A simple
experiment highlighting this behavior consists in spinning an
impeller of radius $R$ and height $h$ at constant angular velocity
$\omega$ inside a tank filled with fluid of density $\rho$ and
kinematic viscosity $\nu$: when the Reynolds number $Re=R^2 \omega
/ \nu$ is large enough, the torque $\Gamma$ required to drive the
impeller follows the $\nu$-independent scaling $\Gamma = K \rho
R^4 h \omega^2$, where the dimensionless drag coefficient $K$
depends only on the shape of the impeller.

Here we consider the effect of global rotation at constant rate
$\Omega$ on this fundamental experiment: how does the drag
coefficient depend on  the Rossby number $Ro=\omega/\Omega$?
Global rotation is encountered in many industrial, geophysical and
astrophysical flows. Rotating turbulence has therefore been
studied intensively using experimental, theoretical and numerical
tools~\citep{DavidsonBook2013,Godeferd2015}. For strong global
rotation, the behavior of rotating turbulent flows may be
summarized as follows: the large-scale flow structures tend to
become invariant along the global rotation axis, in qualitative
agreement with the Taylor-Proudman theorem~\citep{GreenspanBook},
while the remaining vertically dependent fluctuations can be
described in terms of inertial waves that interact nonlinearly,
together and with the 2D
flow~\citep{Clark2014,Yarom2014,Campagne2014,Campagne2015,Alexakis2015}. Rotating turbulence
is therefore intermediate between 2D and 3D turbulence; one
naturally wonders how its energy dissipation rate compares to the
laminar dissipation of 2D turbulence, or to the dissipation
anomaly of 3D turbulence.

The experiment considered here probes directly the influence of
global rotation on turbulent dissipation. Indeed, torque measurements give
access to the drag coefficient:
\begin{equation}
K = \Gamma / (\rho R^4 h \omega^2)  \, , \label{defK}
\end{equation}
i.e., to the normalized energy dissipation rate. We
report on the behavior of $K$ as a function of the Rossby number
$Ro$, in the fully turbulent regime where $K$ is independent of
$Re$, for a rotation axis of the impeller either parallel or
perpendicular to the global rotation axis.

\section{Experimental setup}

\begin{figure}
    \centerline{\includegraphics[width=8cm]{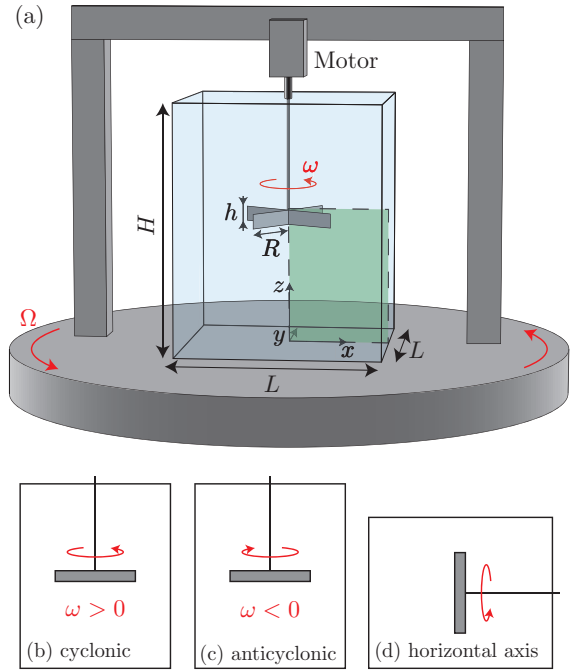}}
    \caption{Experimental setup. We measure the mean torque $\Gamma$ developed
    by the motor driving an impeller at constant rotation rate
    $\omega$ in a water-filled tank mounted on a platform  rotating at
    rate $\Omega$ ($L=45$~cm, $H=55$~cm, $R=12$~cm, $h=3.2$~cm). In
    (a), (b) and (c), the platform and the impeller
    rotate around the same axis (in the laboratory frame, the impeller
    spins at a rate $\omega+ \Omega$). PIV measurements are performed
    in a vertical plane (green dashed region). In (d), the axis
    of the impeller is perpendicular to the global rotation axis of
    the platform.}\label{fig:disp_exp}
\end{figure}

The experimental setup is sketched in
figure~\ref{fig:disp_exp}(a). It consists of a parallelepipedic
water-filled tank of height $H=55$~cm and square base of side
$L=45$~cm. A brushless servo-motor drives a four-rectangular-blade
impeller of radius $R=12$~cm and height $h=3.2$~cm at constant
angular velocity $\omega$ between 20 and 400~rpm. The tank and the
motor are mounted on a 2-meter-diameter platform rotating at
constant rate $\Omega>0$ up to $30$~rpm around the vertical axis.
For each set of parameters $(\omega,\Omega)$, we measure the
time-averaged torque $\Gamma$ developed by the motor driving the
impeller. The maximum applied torque is $\Gamma_m = 5$~N~m. We
subtract the non-hydrodynamic torque, determined for each value of
$\omega$ by repeating the measurement using air instead of water.
This non-hydrodynamic torque includes losses in the motor and in
the o-ring seal through which the shaft enters the tank. It is
determined with a precision of $\Delta \Gamma = 50$~mN~m, allowing
us to span a range of two orders of magnitude in hydrodynamic
torque.

Two configurations are considered. In the first configuration, the
impeller rotates around the vertical axis, either cyclonically
($\omega>0$, figure~\ref{fig:disp_exp}b) or anti-cyclonically
($\omega<0$, figure~\ref{fig:disp_exp}c); in the second
configuration, the impeller rotates around a horizontal axis in
the rotating frame (figure~\ref{fig:disp_exp}d). These
configurations allow us to examine the two physically relevant
situations of a driving velocity either normal or parallel to the
global rotation axis. The vertical-axis configuration
(figure~\ref{fig:disp_exp}b,c) bears some similarities with the
Taylor-Couette flow between rotating cylinders, the key difference
being that the flow is driven inertially: the Taylor-Couette
geometry is well-suited to study the effect of global rotation on
viscous friction near a smooth wall, while our experiment
considers the turbulent drag due to the inertially driven flow.

\section{Drag measurements}

\begin{figure}
    \centerline{\includegraphics[width=9cm]{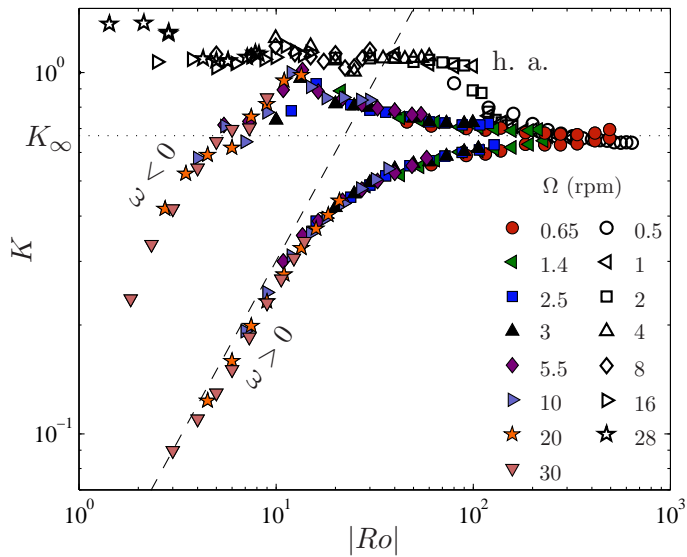}}
    \caption{Drag coefficient $K=\Gamma/\rho R^4 h \omega^2$ as a
    function of the Rossby number $|Ro|=|\omega|/\Omega$. Color
    symbols: vertical-axis configuration, cyclonic and anticyclonic
    (figure~\ref{fig:disp_exp}{b,c}); Open symbols: horizontal-axis
    configuration (figure~\ref{fig:disp_exp}d). The horizontal
    dotted line is the drag coefficient for fully developed turbulence
    without rotation, $K_\infty= 0.67 \pm 0.02$. The tilted dashed
    line shows $K\propto Ro$.}\label{fig:K_Ro}
\end{figure}

We first focus on the vertical-axis configuration.
Figure~\ref{fig:K_Ro} shows the drag coefficient
(\ref{defK}) as a function of the
Rossby number $|Ro|=|\omega|/\Omega$. $K$ is related to the
spatial distribution of energy dissipation rate per unit mass
$\epsilon({\bf x})=\nu \langle | \bnabla {\bf u}|^2  \rangle_{t}$
through the balance between input and dissipated power: $\Gamma
\omega = \rho V \langle\epsilon \rangle_{\bf x}$, where $\langle
\, \rangle_{t}$ denotes a time average and $\langle \,
\rangle_{\bf x}$ the space average over the volume $V=L^2H$ of the
tank. In the absence of global rotation, because of the large
Reynolds number of the flow ($Re = 6.4\times 10^{4}$ to $6.7\times
10^{5}$), the drag coefficient is independent of $Re$, $K_\infty=
0.67 \pm 0.02$, in agreement with the fully turbulent scaling law
$\epsilon \sim R^3 \omega^3/h$. For nonzero global rotation
$\Omega>0$, the drag coefficient remains independent of $Re$, but
it is now a function of the Rossby number: the high-$Re$ data for
$\Gamma(\omega,\Omega)$ collapse onto two master branches when
plotted as $K$ vs. $|Ro|$, a cyclonic branch for $\omega>0$ and an
anticyclonic branch for $\omega <0$.

\begin{figure}
    \centerline{\includegraphics[width=12cm]{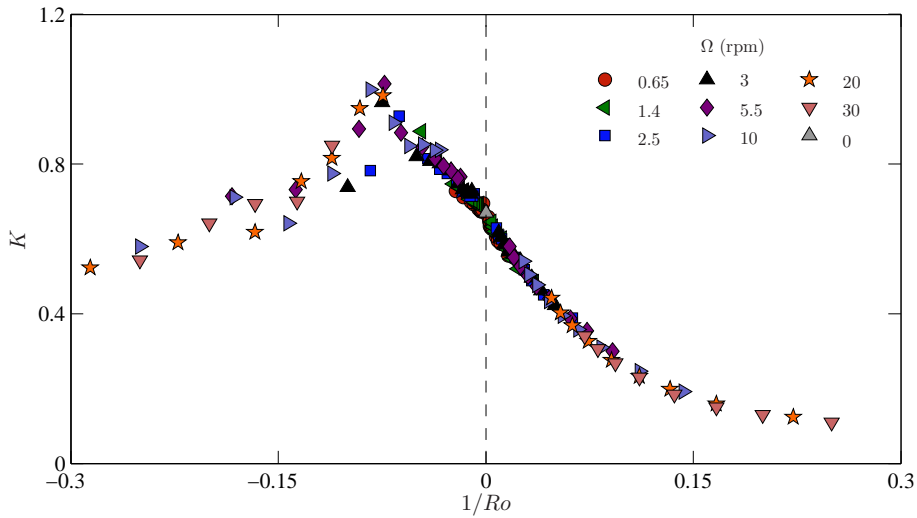}}
    \caption{The drag coefficient $K$ departs from $K_\infty$ approximately linearly in inverse Rossby number $Ro^{-1}=\Omega/\omega$, when the latter is small (same data as figure \ref{fig:K_Ro}).}\label{fig:K_1overRo}
\end{figure}

We first consider the high-$Ro$ part of figure~\ref{fig:K_Ro}. For
weak global rotation (large $|Ro|$), the two branches split
symmetrically about $K_\infty$, with drag reduction for cyclonic
motion ($Ro>0$) and drag enhancement for anticyclonic motion
($Ro<0$). Such small departure from $K_\infty$ can be described
through a regular expansion in $Ro^{-1}=\Omega / \omega$, assuming that the weak Coriolis force can be accounted for perturbatively: writing the
velocity field as ${\bf u}={\bf u}_0+Ro^{-1}{\bf
u}_1+\mathcal{O}(Ro^{-2})$, where ${\bf u}_0$ is the flow
without global rotation, leads to ${\bf u}_1|_{-\Omega}={\bf
u}_1|_\Omega$. The mean energy dissipation rate per unit mass is
$\langle\epsilon \rangle_{\bf x} = \nu \langle | \bnabla {\bf u}_0
|^2  \rangle_{{\bf x},t} + 2 \nu Ro^{-1} \langle \bnabla {\bf
u}_0 \cdot  \bnabla {\bf u}_1 \rangle_{{\bf x},t}
+\mathcal{O}(Ro^{-2})$, where $\langle \, \rangle_{{\bf x},t}$
denotes space and time average, and $\nu \langle | \bnabla {\bf
u}_0 |^2 \rangle_{{\bf x},t}$ is the energy dissipation rate of
the non-rotating flow. The drag coefficient becomes
\begin{equation}\label{eq:Kexpanded}
\frac{K}{K_\infty}=1+\alpha Ro^{-1} + \mathcal{O}(Ro^{-2}) \, , \mbox{ where } \alpha = 2  \frac{\langle \bnabla {\bf
u}_0 \cdot  \bnabla {\bf u}_1  \rangle_{{\bf
x},t}}{\langle | \bnabla {\bf u}_0 |^2  \rangle_{{\bf x},t}} \, .
\end{equation}
The sign of $\alpha$ can be inferred from the
fact that global rotation tends to reduce the velocity gradients
along the rotation axis. In the laboratory frame, and for fixed
$\Omega>0$, the fluid spins faster for cyclonic rotation of the
impeller ($\omega>0$) than for anticyclonic rotation ($\omega<0$).
As a consequence, we expect lower vertical velocity gradients in
the cyclonic case, i.e., $\alpha < 0$, implying consistently a
negative correlation between the perturbed vertical derivative
$\partial_z {\bf u}_1$ and $\partial_z {\bf u}_0$ in
equation~(\ref{eq:Kexpanded}). The data in
figure~\ref{fig:K_1overRo} are in good agreement with this
prediction: the departure of $K$ from $K_\infty$ is approximately linear in $Ro^{-1}$ for $Ro^{-1} \in [-0.07; 0.07]$, which corresponds to the range $|Ro| > 15$ in figure~\ref{fig:K_Ro}.

We now discuss the regime $Ro \simeq 1$, which is probably the
most interesting one: the cyclonic branch of figure~\ref{fig:K_Ro}
displays a dramatic drop in drag coefficient, with $K$ reaching
values as low as $12$\% of $K_\infty$ for the lowest Rossby
numbers. This decrease in drag with increasing $\Omega$ follows
the approximate scaling law $K \sim Ro$. A similar -- although
less pronounced -- decrease in drag coefficient is observed for
the anticyclonic branch. However, it is preceded by a dissipation
peak at intermediate $Ro$: a maximum dimensionless drag
$K_{\text{peak}} \simeq 1.5 K_\infty$ is achieved for $Ro_{peak}
\simeq -12$. Once again, this difference between the two branches
can be related to the effective global rotation of the fluid: the
dissipation peak corresponds to anticyclonic impeller motion
$\omega<0$ partly compensating the rotation $\Omega>0$ of the
platform, so that the fluid has minimum global rotation in the
laboratory frame. A similar peak of dissipation has been reported
for counter-rotating Taylor-Couette flow, and corresponds to
optimal transport of angular momentum between the two
cylinders~\citep{Dubrulle2005,Gils2011,Paoletti2011,Ostilla2014}.
Such intermediate negative Rossby numbers also represent the most
unstable configuration for vortices in rotating
flows~\citep{Kloosterziel1991,Mutabazi1992}, which rapidly break
down into highly-dissipative 3D structures. The Taylor-Couette
dissipation peak can then be traced back to strong instabilities
driving dissipative 3D flow structures. In a similar fashion, the
PIV measurements described in section \ref{secPIV} indicate that
highly-dissipative 3D flow structures are responsible for the
dissipation peak of the present experiment (see figure
\ref{fig:PIV}c).

How to explain such a strong drag reduction for rapid global
rotation? Two scenarios can be put forward. A first scenario
relies on the modification of the energy transfers by the
background rotation. In this approach, the velocity fluctuations
are described in terms of propagating inertial waves, which
disrupt the phase relation needed for efficient energy
transfers~\citep{Cambon1989}. This scenario was first put forward
in the context of magnetohydrodynamic
turbulence~\citep{Iroshnikov,Kraichnan1965}, and later applied to
rapidly rotating turbulence~\citep{Zhou1995,Smith1996}. It
predicts reduced energy transfers $\epsilon \simeq \epsilon_\infty
Ro'$ in the limit $Ro' \ll 1$, where $\epsilon_\infty \simeq
u'^3/\ell$ is the usual (non-rotating) dissipation rate
constructed on the turbulent velocity $u'$ and the
energy-containing size $\ell$, and $Ro' = u' / 2 \Omega \ell$ is the
turbulent Rossby number. This result, which relies on dimensional
analysis, can be made more quantitative in the framework of wave
turbulence theory \citep{Galtier2003,Cambon2004,Nazarenko2011}.
Even though the latter theory is valid for $Ro' \ll 1$ only (to be compared to the lowest value $Ro' \simeq 0.2$ of the present study, see Table \ref{table:param}), the scaling law $K \sim Ro$ reported in figure~\ref{fig:K_Ro} turns
out to be in agreement with this prediction, if one assumes that
$u' \sim R \omega$ holds regardless of $Ro$.

Another explanation for the decrease in drag coefficient is partial
two-dimensionalization. Indeed, the forcing geometry in the
vertical-axis configuration (figure~\ref{fig:disp_exp}b,c)  is
compatible with the Taylor-Proudman theorem, which predicts for
low $Ro$ a purely 2D vertically invariant solution corresponding
to solid-body rotation in the cylinder tangent to the impeller, at
angular velocity $\omega$ in the rotating frame of the platform.
This solution is valid for a perfect fluid only, which slips on
the top and bottom boundaries. For a realistic viscous fluid,
additional boundary layers and poloidal recirculations develop and
coexist with the solid-body motion~\citep{Hide,GreenspanBook}. As
a matter of fact, for periodic boundary conditions and idealized
vertically invariant forcing, it was recently proven that the
high-$Re$ flow settles in a purely 2D state for low enough Rossby
number~\citep{Gallet}. For the experiment at stake here, the
rapidly rotating flow can be thought of as a coexistence of this
2D asymptotic flow, together with weak 3D turbulent fluctuations,
the intensity of which decreases as global rotation increases. The
2D flow dissipates very little energy, at a laminar rate,
typically proportional to viscosity. Accordingly, energy
dissipation is due to the 3D flow structures,
the intensity of which decreases for decreasing $Ro$. We therefore
expect a decrease in energy dissipation -- and therefore in drag
coefficient -- for increasing global rotation rate.

\section{Turbulent flow structure\label{secPIV}}

\begin{figure}
    \centerline{\includegraphics[width=12cm]{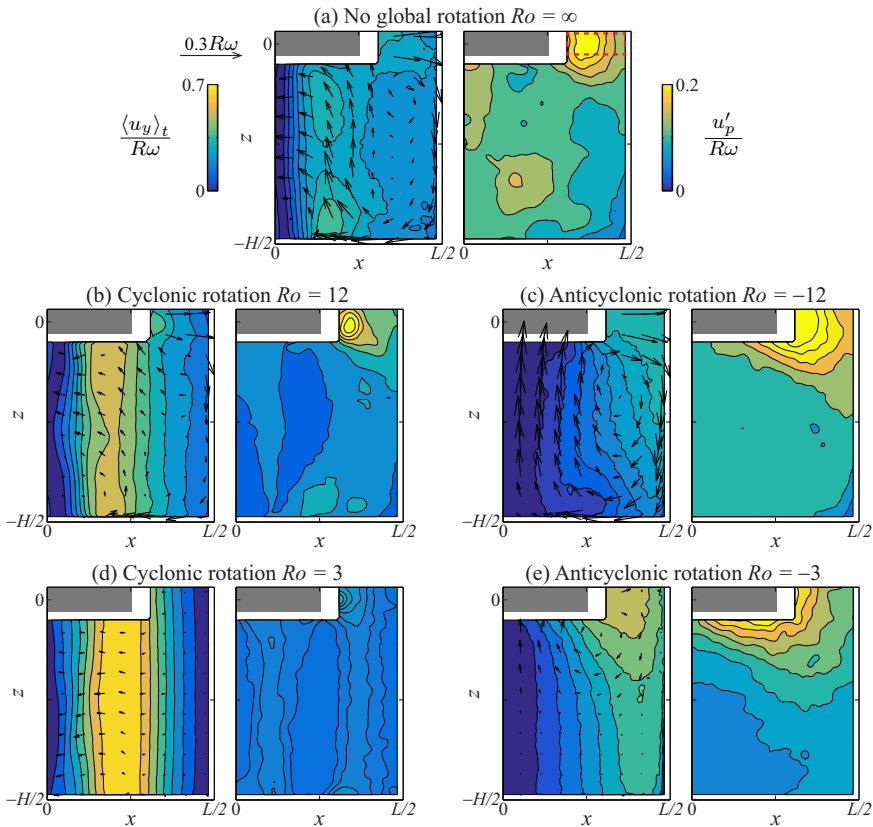}}
    \caption{Time-averaged flow (left) and rms fluctuations of the
    poloidal flow (right), measured by stereoscopic PIV in the frame
    of the rotating platform. In the left panels, the color codes the
    toroidal (out-of-plane) velocity. (a), no background
    rotation ($\omega=150$~rpm, $\Omega=0$). (b,d), cyclonic
    background rotation ($Ro=12$ and 3). The two-dimensionalization
    results in a gradual weakening of the poloidal flow ${\bf u_p}$
    and a strengthening of the toroidal flow; at the largest global
    rotation (d), the fluid column below the impeller rotates
    rigidly at $\omega$ in the rotating frame, with weak turbulent
    fluctuations. (c,e), anticyclonic background rotation
    ($Ro=-12$ and $-3$). The peak dissipation at $Ro\simeq-12$ in
    figure~\ref{fig:K_Ro} corresponds to the maximum poloidal
    recirculation and maximum turbulent fluctuations in the vicinity
    of the impeller (panel c).}\label{fig:PIV}
\end{figure}

\begin{table}
\centering
\begin{tabular}{c|c|c|c|c}
Panel & $Ro=\dfrac{\omega}{\Omega}$ & $Re=\dfrac{\omega R^2}{\nu}$ & $Ro'=\dfrac{u'_p}{2\Omega h}$ & $Re'=\dfrac{u'_p h}{\nu}$ \\ \hline
(a) & $\infty$ & $2.3\times10^5$ & $\infty$ & $7.2\times10^3$\\
(b) & $12$ & $1.8\times10^5$ & $2.4$ & $5.1\times10^3$\\
(c) & $-12$ & $1.8\times10^5$ & $3.3$ & $7.2\times10^3$\\
(d) & $3$ & $1.4\times10^5$ & $0.2$ & $1.6\times10^3$\\
(e) & $-3$ & $1.4\times10^5$ & $0.7$ & $4.4\times10^3$\\
\end{tabular}
 \caption{Dimensionless numbers for the five panels of figure~\ref{fig:PIV}, based either on the control parameters or on the fluctuating poloidal velocity (inside the red-dashed domain of figure~\ref{fig:PIV}a).}\label{table:param}
\end{table}

To discriminate between the inertial-wave scenario and the partial
two-dimensionalization one, we measure the velocity field using a
stereoscopic PIV system mounted on the rotating platform. The
field of view is vertical, illuminated by a laser sheet containing
the axis of the impeller, and represents one quarter of the tank
section, below the impeller (see figure~\ref{fig:disp_exp}a).
Stereoscopic PIV measurements are achieved by two high-resolution
cameras aimed at the measurement plane at an incidence angle of
45$^\circ$, at a frame rate of 5~Hz.

In figure~\ref{fig:PIV}, we show the mean velocity field, and the
standard deviation $u'_p(x,z)$ of the poloidal velocity, with and
without global rotation. The non-rotating mean flow corresponds to
a toroidal (out-of-plane) flow driven by the propeller, together
with a strong poloidal (in-plane) recirculation. The flow displays
3D turbulence, the intensity of which is maximum at the edge of
the impeller's blades. In table \ref{table:param}, we provide values of the turbulent Reynolds and Rossby numbers based on the rms poloidal velocity in this flow region (red-dashed domain in figure \ref{fig:PIV}a).

\begin{figure}
    \centerline{\includegraphics[width=8cm]{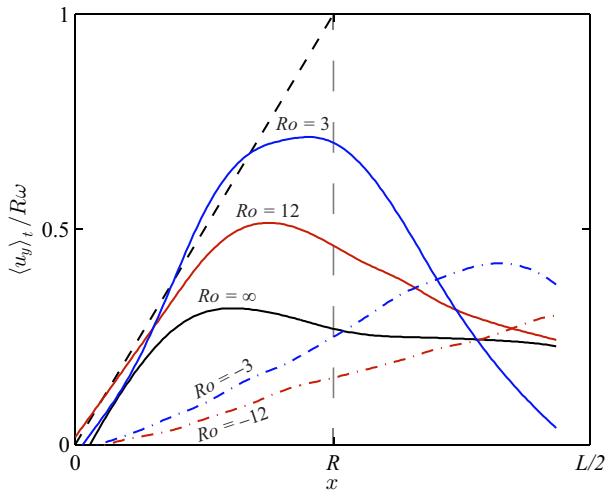}}
    \caption{Mean azimuthal (out-of-plane) velocity profile $\langle
    u_y \rangle_t$, measured 200 mm below the propeller ($z/H = -
    0.36$), showing the strong two-dimensionalization of the flow for
    cyclonic background rotation ($Ro>0$). The dashed line shows the
    Taylor-Proudman prediction $\langle u_y \rangle_t = \omega x$
    (solid-body rotation at angular velocity $\omega$ in the rotating
    frame).}\label{fig:profile}
\end{figure}

The non-rotating flow contrasts strongly with the rapidly
rotating one measured for cyclonic impeller motion: in line with
the Taylor-Proudman theorem, the mean toroidal flow gradually
tends to solid-body rotation at frequency $\omega$ inside the
tangential cylinder, while the mean poloidal recirculation weakens
as $\Omega$ increases. This two-dimensionalization is clearly
visible in figure~\ref{fig:profile}, which shows the mean
azimuthal velocity profile well below the impeller ($z/H=-0.36$).
Another consequence of the Taylor-Proudman theorem is a strong
decrease of the 3D turbulent fluctuations for decreasing Rossby
number, as can be seen on the right-hand panels of
figure~\ref{fig:PIV}b,d: in the vicinity of the blades, i.e., inside the red-dashed domain sketched in figure~\ref{fig:PIV}a, both the poloidal and toroidal rms velocities strongly decrease for decreasing Rossby number (we show only the former), the ratio of the two being approximately $1.2$ regardless of $Ro$.

The anticyclonic case is different: for intermediate global
rotation ($Ro=-12$, figure~\ref{fig:PIV}c), the mean flow displays
minimum toroidal component and maximum poloidal recirculation, and
the 3D turbulent structures in the vicinity of the propeller have
maximum intensity. This situation corresponds to the dissipation
peak in figure~\ref{fig:K_Ro}. However, as global rotation is
further increased ($Ro=-3$, figure~\ref{fig:PIV}e), the measured
flows once again follow the Taylor-Proudman phenomenology, with
stronger toroidal mean flow and reduced mean and fluctuating
poloidal velocities (although the flow remains further from the 2D
state than for cyclonic propeller motion).

\begin{figure}
    \centerline{\includegraphics[width=8cm]{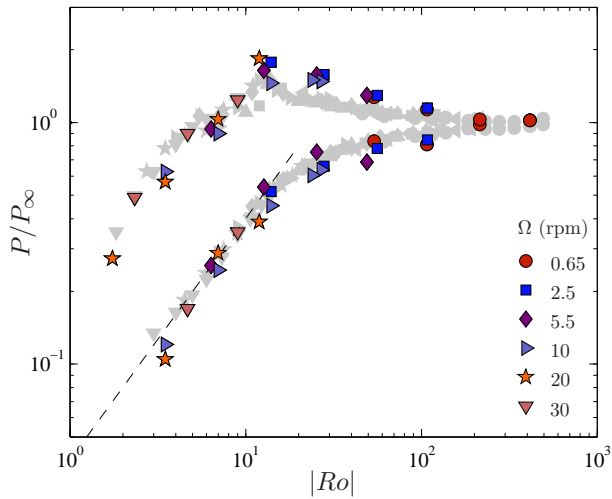}}
    \caption{Dissipated power $P$, estimated by the turbulent
    fluctuations $u_p'$ of the poloidal flow measured through
    stereoscopic PIV, using the non-rotating estimate $u_p^{\prime
    3}/h$, and normalized by its non-rotating value $P_\infty$. The
    good agreement between $P/P_\infty$ and the input power
    $K/K_\infty$ measured from torque data (data shown as faint
    symbols reproduced from figure~\ref{fig:K_Ro}) demonstrates that the
    non-rotating estimate $u_p^{\prime 3}/h$ correctly describes the
    energy dissipation rate, even in the rotating
    case.}\label{fig:dissipation}
\end{figure}

On a qualitative level, these observations therefore support the
partial two-dimensionalization scenario: most of the kinetic
energy of the rapidly-rotating flow corresponds to 2D
$z$-invariant motion, which is discarded at the outset of wave
turbulence theories. A quantitative criterion is however needed to
distinguish more clearly between the two scenarios. To wit, we
evaluate the energy dissipation rate of the turbulent poloidal
flow directly from the PIV measurements: assuming that the
energy-containing scale is given by the impeller height $h$, the
non-rotating estimate of this quantity is $u_p^{\prime 3}/h$,
whereas the wave turbulence estimate is $u_p^{\prime 4}/\Omega
h^2$. We compute the spatial integral of the non-rotating estimate
$u_p^{\prime 3}/h$ in the vicinity of the impeller, inside the
dashed region shown in the upper-right panel of
figure~\ref{fig:PIV}. We denote as $P$ the resulting quantity, and
report its behavior with $Ro$ in figure~\ref{fig:dissipation}. It
matches remarkably the behavior of $K$, both curves being
normalized by their asymptotic non-rotating value, with $K_\infty
/ P_\infty = 0.6 \pm 0.1$. This good agreement confirms that
energy dissipation is mostly due to the 3D part of the flow, the
laminar dissipation of the 2D flow being negligible. But more
importantly,  for the moderate Rossby numbers studied here, it
demonstrates that energy dissipation can be estimated locally from
the usual non-rotating scaling-law $u_p^{\prime 3}/h$, ruling out
the modified energy dissipation $u_p^{\prime 4}/\Omega h^2$
predicted by weak turbulence theory (such non-rotating scaling-law is also consistent with studies of grid-generated rotating turbulence \citep{Hopfinger1982,Staplehurst2008,Moisy2011}, which showed that the small-scale turbulent fluctuations start developing some rotation-induced anisotropy for $Ro' \lesssim 0.2$ only). The quantitative criterion illustrated in figure~\ref{fig:dissipation}
allows to clearly discriminate between the two scenarios, and we
conclude that drag reduction originates from a partial
two-dimensionalization of the flow, with reduced 3D turbulent
fluctuations.

\section{Concluding remarks}

The two-dimensionalization scenario for drag reduction is
supported so far by the vertical axis configuration
(figure~\ref{fig:disp_exp}b,c), which is compatible with the
Taylor-Proudman theorem. What if the axis of the impeller is
horizontal (see figure~\ref{fig:disp_exp}d)? Such a configuration
is incompatible with the Taylor-Proudman theorem: there is no
vertically invariant flow solution compatible with the boundary
conditions, even for a perfect fluid, because of the nonzero
vertical velocity imposed by the impeller. This geometry therefore
prohibits the partial two-dimensionalization scenario, while it
still allows for the inertial-wave one. The vertical motion of the
blades induces 3D turbulent velocity fluctuations of order $R
\omega$, regardless of $Ro$. The usual non-rotating estimate for
the energy dissipation rate then gives $R^3 \omega^3/h$, from
which we predict no dependence of $K$ on $Ro$, whereas the wave
turbulence prediction gives $R^4 \omega^4/h^2 \Omega$, with a
strong decrease in drag coefficient $K\sim Ro$ for decreasing
$Ro$. Once again, the data in figure~2  clearly depart from
the wave-turbulence prediction: for the moderate Rossby numbers
considered here, there is no drop in drag coefficient in this
horizontal-axis configuration.

To conclude, we observe strong drag reduction whenever
two-dimensionalization is allowed, i.e., when the forcing geometry
is compatible with the Taylor-Proudman theorem:  the drag
coefficient is dramatically reduced for motion perpendicular to
the global rotation axis, while it is very weakly affected for
motion parallel to the global rotation axis. Importantly, for the
moderate Rossby numbers considered here, the energy dissipation
rate obeys the classical non-rotating scaling-law, and not the
wave-turbulence one. The decrease in drag is a consequence of a
decrease in 3D turbulence intensity, in line with the
two-dimensionalization of the large-scale flow. It would be of
great interest to achieve even faster global rotation, to determine how the flow approaches the asymptotic 2D state: is there a threshold $Ro$ under which the flow becomes exactly 2D (up to boundary layers), as in the stress-free case considered by \citet{Gallet}, or is there a clear-cut scaling-law governing the decrease in 3D energy as a function of $Ro$?

Such experiments would also indicate how much further the drag can be reduced: for very fast global
rotation, the decrease in bulk dissipation may be hidden by
increasing Ekman friction in the boundary layers, which probably
becomes the dominant cause of dissipation for large $\Omega$. As a
consequence, there would be an optimum rotation rate that leads to
a minimum in drag.

We acknowledge S. Fauve for providing the fruitful seed idea, and  A. Aubertin, L. Auffray, C. Borget and R. Pidoux
for their experimental help. This work is supported by
``Investissements d'Avenir'' LabEx PALM (ANR-10-LABX-0039-PALM).
F.M. acknowledges the Institut Universitaire de France.

\end{document}